\documentclass[12pt,preprint]{elsarticle} \usepackage{graphicx}
\usepackage{amssymb}
\usepackage{amsmath}

\def\bea{\begin{eqnarray}}
\def\eea{\end{eqnarray}}
\def\beq{\begin{eqnarray}}
\def\eeq{\end{eqnarray}}

\def\bm{\begin{math}}
\def\me{\end{math}}

\begin{document}
\begin{frontmatter}
\title{ Domain Growth in Chiral Phase Transitions}
\author[jnu]{Awaneesh Singh}
\ead{awaneesh11@gmail.com}
\author[jnu]{ Sanjay Puri}
\ead{purijnu@gmail.com}
\author[prl,jnu]{Hiranmaya Mishra \corref{cor}}
\ead{hm@prl.res.in}
\cortext[cor]{Corresponding author}
\address[jnu]{School of Physical Sciences, Jawaharlal Nehru University, New Delhi, 110067,
India}
\address[prl]
 {Theory Division, Physical Research Laboratory, Navrangpura, Ahmedabad -- 380009, India.}
\begin{abstract}
We study the kinetics of chiral phase transitions in quark matter. We discuss the phase diagram of this system in both a microscopic framework (using the Nambu-Jona-Lasinio model) and a phenomenological framework (using a Landau free energy). Then, we study the far-from-equilibrium coarsening dynamics subsequent to a quench from the chirally-symmetric phase to the massive quark phase. Depending on the nature of the quench, the system evolves via either {\it spinodal decomposition} or {\it nucleation and growth}. The morphology of the ordering system is characterized using the order-parameter correlation function, structure factor, domain growth laws, etc.
\end{abstract}
\begin{keyword}{Chiral symmetry breaking, Ginzburg-Landau expansion, TDGL equation, domain growth, quenching, dynamical scaling.}
\end{keyword}
\end{frontmatter}

\section{Introduction}
\label{intro}

The nature of the {\it quantum chromodynamics} (QCD) phase diagram as a 
function of temperature ($T$) and baryon chemical potential ($\mu$) has been 
studied extensively over the last few years \cite{rischke}. For $\mu=0$,
 finite-temperature calculations have been complemented by first-principle approaches 
like lattice QCD simulations \cite{karsch}. However, for $\mu \ne 0$, the lattice 
simulations are limited to small values of $\mu$ \cite{latmu}. In QCD with two 
massless quarks, the chiral phase transition is expected to be second-order at 
zero baryon densities. In nature, the light quarks are not exactly massless and 
the sharp second-order transition is replaced by a smooth crossover. This picture 
is consistent with lattice QCD simulations with a transition temperature $T_c\sim 140 
- 190$ MeV \cite{Tclattice}. On the other hand, calculations based on different 
effective models show that the transition becomes first-order at large $\mu$ and
 small $T$ \cite{mk98}. This means that the phase diagram will have 
a {\it tricritical point} (TCP), where the first-order chiral transition 
becomes second-order (for vanishing quark masses) or ends (for non-vanishing 
quark masses). The location of the TCP ($\mu_{\rm tcp}$, $T_{\rm tcp}$) in the 
phase diagram and its signature has been been under intense investigation, both theoretically and experimentally \cite{gavaigupta,misha}.

Heavy-ion collision experiments at high energies produce hot and dense strongly-interacting
matter, and provide the opportunity to explore the phase diagram of QCD. The high-$T$ and
low-$\mu$ region has been explored by recent experiments in the
{\it Relativistic Heavy Ion Collider} (RHIC). This region will also be studied by planned
experiments in the {\it Large Hadron Collider} (LHC). Further, future heavy-ion collision
experiments like the {\it Beam Energy Scan} at RHIC, {\it FAIR} in GSI, and {\it NICA} in
Dubna plan to explore the high-baryon-density region of the phase diagram, particularly
around the TCP \cite{cpod}. The experiments at the RHIC provide clear signals that nuclear
matter undergoes a phase transition to partonic phases at sufficiently large values of the
energy density. However, the nature and kinetics of this transition remains an open question.
We note here that lattice QCD assumes that the system is in equilibrium, whereas
heavy-ion experiments are essentially nonequilibrium processes. Therefore, information about
which equilibrium phase has the lowest free energy is not sufficient to discuss the properties
of the system. One also has to understand the kinetic processes which
drive the phase transition, and the properties of the nonequilibrium structures that the
system must go through to reach equilibrium.

In this paper, we study the kinetics of chiral transitions in quark matter. We focus on far-from-equilibrium kinetics, subsequent to a quench from the disordered phase (with zero quark condensate) to the ordered phase. This rapid quench renders the disordered system thermodynamically unstable. The evolution to the new equilibrium state is characterized by spatio-temporal pattern formation, with the emergence and growth of domains of the preferred phases. This nonlinear evolution is usually referred to as {\it phase ordering dynamics} or {\it coarsening} or {\it domain growth} \cite{aj94,pw09}. Previous studies of ordering dynamics in quark matter, which we review shortly, have primarily focused upon early-time kinetics and the growth of initial fluctuations. The present paper is complementary to these studies. We investigate the universal properties (e.g., growth laws, scaling of correlation and structure functions, bubble dynamics, etc.) in the late stages of chiral kinetics. These properties are robust functions of the evolution dynamics, and only depend upon general features, e.g., scalar vs. vector order parameters, defect structures, conservation laws which govern dynamics, relevance of hydrodynamic effects, etc. Our results in this paper are obtained using a {\it time-dependent Ginzburg-Landau} (TDGL) model, which is derived from the {\it Nambu-Jona-Lasinio} (NJL) model with two light flavors \cite{klevansky}. However, we expect that our results apply to a much larger class of systems belonging to the same {\it dynamical universality class}.

To place our work in the proper context, we provide an overview of studies of dynamical properties of quark matter. These focus on either (a) {\it critical dynamics}, i.e., time-dependent behavior in the vicinity of the critical points, or (b) {\it far-from-equilibrium dynamics}, which was explained above. In studies of critical dynamics, much interest has focused on the TCP. The {\it static universality class} of the QCD transition (for non-vanishing quark masses) is believed to be that of the $d=3$ Ising model, but there is debate regarding the dynamical universality class. For example, the TCP dynamics was argued \cite{bednikov} to be in the class of {\it Model C} in the Hohenberg-Halperin classification scheme \cite{hohenrev}. Essentially, the argument was that critical dynamics is described by a nonconserved order parameter (the quark condensate $\langle\bar\psi\psi\rangle$), in conjunction with conserved quantities like the baryon number density. If one includes the mode coupling between the quark condensate and the baryon density, the appropriate universality class is {\it Model H} with a different dynamical exponent \cite{son}. However, it was also argued that reversible couplings can play a crucial role in QCD critical dynamics, which may then differ from that of {\it Model H} \cite{koide}. In related work, Koide and Maruyama \cite{tomoinpa} have derived a linear Langevin equation for the chiral order parameter. This is obtained by applying the {\it Mori projection operator technique} to the NJL model. These authors study the solution of the Langevin equation, and investigate {\it critical slowing down} in the vicinity of the TCP.

Let us next turn our attention to studies of far-from-equilibrium dynamics in quark matter. Sasaki et al. \cite{sasaki} have emphasized the importance of {\it spinodal decomposition} in understanding the chiral and deconfinement transitions in heavy-ion collisions. Their study was based on a mean-field approximation to the NJL model, as well as a phenomenological Landau theory. Sasaki et al. discussed fluctuations of the baryon number density as possible signatures of nonequilibrium transitions. However, they did not study the corresponding evolution dynamics. Scavenius et al. \cite{dumitru} have investigated the possibility of {\it nucleation} vs. {\it spinodal decomposition} in an effective field theory derived from the {\it nonlinear sigma model}. Again, these authors have not investigated time-dependent properties, which are of primary interest to us in this paper.

An important study of evolution dynamics in {\it quark-gluon plasma} (QGP) is due to Fraga and Krein \cite{fragaplb}, who modeled the relaxation to equilibrium via a phenomenological Langevin equation. (We follow a similar approach in this paper.) This Langevin equation can be derived from a microscopic field-theoretic model of kinetics of the chiral order parameter \cite{gr94,dr98}. Fraga and Krein studied the early-time dynamics of spinodal decomposition in this model both analytically and numerically, and focused upon the effect of dissipation on the spinodal instability. In recent work, Bessa et al. \cite{fraga} studied bubble nucleation kinetics in chiral transitions, and the dependence of the nucleation rate on various parameters.

Skokov and Voskresensky \cite{skokovdima} have also studied the kinetics of first-order phase transitions in nuclear systems and QGP. Starting from the equations of non-ideal non-relativistic hydrodynamics (i.e., Navier-Stokes equation, continuity and transport equations), they derived TDGL equations for the coupled order parameters. These TDGL equations were studied numerically and analytically in the vicinity of the critical point. Skokov-Voskresensky focus upon the evolution of density fluctuations in the metastable and unstable regions of the phase diagram, and the growth kinetics of seeds. They clarify the role of viscosity in the ordering kinetics. Finally, we mention the recent work of Randrup \cite{jr09}, who has studied the fluid dynamics of relativistic nuclear collisions. The corresponding evolution equations reflect the conservation of baryon charge, momentum and energy. Randrup studied the amplification of spinodal fluctuations and the evolution of the {\it real-space correlation function} and the {\it momentum-space structure factor}. (We will study the scaling of these quantities in Sec.~\ref{kct} of this paper.) Randrup's work mostly focused upon the evolution in the linearized regime, where there is an exponential growth of initial fluctuations.

This paper is complementary to Refs.~\cite{fragaplb,fraga,skokovdima,jr09}, and investigates the late stages of phase-separation kinetics in quark matter. The system is described by nonlinear evolution equations in this regime: the exponential growth of initial fluctuations is saturated by the nonlinearity. We study the coarsening dynamics from disordered initial conditions, and the scaling properties of emergent morphologies. We consider an initially disordered system which is quenched to the symmetry-broken phase through either the second-order line (relevant for high $T$ and small $\mu$) or the first-order line (relevant for small $T$ and large $\mu$) in the NJL phase diagram. We study domain growth for both types of quenching, and highlight quantitative features of the coarsening morphology. We also study the evolution kinetics of single droplets, and the dependence of the front velocity on system parameters.

This paper is organized as follows. In Sec.~\ref{pdct}, we discuss the equilibrium phase diagram of the two-flavor NJL model using a variational approach. Here, we describe chiral symmetry breaking as a vacuum realignment with quark-antiquark condensates. As we shall see, this method also captures some extra contributions proportional to 1/$N_c$ (where $N_c$ is the number of colors), as compared to mean-field theory. In Sec.~\ref{pdct}, we will also discuss the corresponding Landau description of chiral transitions. In Sec.~\ref{kct}, we introduce the TDGL equation which describes the evolution of the chiral order parameter, and use it to study the kinetics of chiral transitions. As mentioned earlier, we focus on pattern formation in the late-stage dynamics, which is characterized by scaling of the evolution morphologies, and the corresponding domain growth laws. Finally, we end this paper with a summary and discussion in Sec.~\ref{summary}. 

Our investigation has several novel features from the perspective of both QCD and domain growth studies. These can be highlighted as follows. First, we demonstrate a quantitative mapping between the phase diagrams of the NJL model as an effective model of QCD at low energy and the $\psi^6$-Landau potential. This mapping enables us to identify the relevant time-scales and length-scales in chiral transition kinetics. Second, we clarify the quantitative features of the coarsening morphology, e.g., correlation functions, growth laws, etc., in chiral transitions. These universal features are independent of system and model details, and can be measured in experiments on quark-gluon plasma. Third, the chiral transition provides a natural context to study ordering dynamics in the $\psi^6$-potential, which has received little attention. To date, most studies of domain growth have focused on the $\psi^4$-potential, which has a much simpler phase diagram. Finally, as we will discuss elsewhere, chiral dynamics also provides a framework to study the effect of inertial terms in phase ordering kinetics. Studies of domain growth have almost entirely focused on dissipative overdamped dynamics \cite{aj94,pw09}.

\section{ Phase Diagram for Chiral Transitions}
\label{pdct}

\subsection{ An Ansatz for the Ground State}
\label{gs}

For the consideration of chiral symmetry breaking, we denote the perturbative vacuum state with chiral symmetry as $|0\rangle$. We then assume a specific vacuum realignment which breaks chiral symmetry because of interactions. Let us first note the quark-field operator expansion in momentum space \cite{hmnj}:
\begin{eqnarray}
\psi (\vec{x}) &\equiv& \frac{1}{(2\pi)^{3/2}}\int \!d\vec{k}\; e^{i\vec{k}\cdot\vec{x}} \tilde\psi(\vec{k}) \nonumber\\ 
&=&\frac{1}{(2\pi)^{3/2}}\int\! d\vec{k} e^{i\vec{k}\cdot \vec{x}} \big[U_0(\vec{k})q^0_I(\vec{k})+V_0(-\vec{k})\tilde q^0_I(-\vec{k} )\big] ,
\label{psiexp}
\end{eqnarray}
where 
\begin{eqnarray}
U_0(\vec{k})=\left(\begin{array}{c}\cos \left(\phi^0/2 \right) \\
\vec{\sigma} \cdot \hat{k} \sin \left(\phi^0/2 \right) \end{array}\right),\;\;
V_0(-\vec{k} )=
\left(\begin{array}{c} - \vec{\sigma} \cdot \hat{k} \sin \left(\phi^0/2 \right) \\ 
\cos \left(\phi^0/2 \right) \end{array}\right).
\label{uv0}
\end{eqnarray}
The superscript $0$ indicates that $q_I^0$ and $\tilde q_I^0$ are two-component operators which annihilate or create quanta, and act upon the chiral vacuum $|0\rangle$. We have suppressed here the color and flavor indices of the quark-field operators. The function $\phi^0(\vec{k})$ in the spinors of Eq.~(\ref{uv0}) is obtained as $\cot{\phi_i^0}(\vec{k}) = m_i/k$ for free massive fermion fields, $i$ being the flavor index. For massless fields, $\phi^0(\vec{k})=\pi/2$.

We now consider vacuum destabilization leading to chiral symmetry breaking \cite{hmnj}, described by
\begin{equation} 
|{\rm vac} \rangle={\cal U}_Q|0\rangle,
\label{u0}
\end{equation} 
where
\begin{equation}
{\cal U}_Q=\exp\left[
\int d\vec{k}~ q_I^{0i}(\vec{k})^\dagger(\vec{\sigma}\cdot\vec{k})h_i(\vec{k})\tilde q_I^{0i} (-\vec{k})-\mathrm{h.c.}\right].
\label{uq}
\end{equation}
Here, $h_i(\vec{k})$ is a real function of $|\vec{k}|~(=k)$ which describes vacuum realignment for quarks of a given flavor $i$. We shall take the condensate function $h_i(\vec{k})$ to be the same ($h_i = h$) for $u$ and $d$ quarks. Clearly, a nontrivial $h_i(\vec{k})$ will break chiral symmetry. A sum over the three colors and three flavors is understood in the exponent of ${\cal U}_Q$ in Eq.~(\ref{uq}).

Finally, to include the effect of temperature and density, we write down the state at nonzero temperature and chemical potential $|\Omega(\beta,\mu)\rangle$, where $\beta = 1/T$. This is done through a thermal Bogoliubov transformation of the state $|\Omega\rangle$, using {\it thermo-field dynamics} (TFD) \cite{tfd,amph4}. We then have
\begin{equation} 
|\Omega(\beta,\mu)\rangle={\cal U}_{\beta,\mu}|\Omega\rangle={\cal U}_{\beta,\mu}{\cal U}_Q |0\rangle,
\label{ubt}
\end{equation} 
where ${\cal U}_{\beta,\mu}$ is
\begin{equation}
{\cal U}_{\beta,\mu}=e^{{\cal B}^{\dagger}(\beta,\mu)-{\cal B}(\beta,\mu)}.
\label{ubm}
\end{equation}
Here, 
\begin{equation}
{\cal B}^\dagger(\beta,\mu)=\int d\vec{k}~ \Big [
q_I^\prime (\vec{k})^\dagger \theta_-(\vec{k}, \beta,\mu)
\underline q_I^{\prime} (\vec{k})^\dagger +
\tilde q_I^\prime (\vec{k}) \theta_+(\vec{k}, \beta,\mu)
\underline { \tilde q}_I^{\prime} (\vec{k})\Big ].
\label{bth}
\end{equation}
In Eq.~(\ref{bth}), the ansatz functions $\theta_{\pm}(\vec{k},\beta,\mu)$ will be related to quark and anti-quark distributions. The underlined operators are defined in the extended Hilbert space associated with thermal doubling in the TFD method. In Eq.~(\ref{bth}), we have suppressed the color and flavor indices on the quarks and the functions $\theta_\pm (\vec{k},\beta,\mu)$. The ansatz functions $h(\vec{k})$, $\theta_\pm(\vec{k},\beta,\mu)$ will be determined by minimizing the thermodynamic potential in the next subsection.

\subsection{Minimization of Thermodynamic Potential and Gap Equations}
\label{gap}

We next consider the NJL model, which is based on relativistic fermions interacting through local current-current couplings. It is assumed that gluonic degrees of freedom can be frozen into point-like effective interactions between the quarks. We shall confine ourselves to the two-flavor case only, with the Hamiltonian
\begin{equation}
{\cal H} = \sum_{i,a}\psi^{ia \dagger}\left(-i\vec{\alpha}\cdot\vec{\nabla} + \gamma^0 m_i \right)\psi^{ia} -G\left[(\bar\psi\psi)^2-(\bar\psi\gamma^5 \tau \psi)^2\right].
\label{ham}
\end{equation}
Here, $m_i$ is the current quark mass. We take this to be the same ($m_i=m$) for both $u$ and $d$ quarks. The parameter $G$ denotes the quark-quark interaction strength. Further, $\tau$ is the Pauli matrix acting in flavor space. The quark operator $\psi$ has two indices $i$ and $a$, denoting the flavor and color indices, respectively. The point interaction produces short-distance singularities and, to regulate the integrals, we restrict the phase space to lie inside the sphere $k< \Lambda$, the ultraviolet cut-off in the NJL model.

We next obtain the expectation values of various operators for the variational ansatz state in Eq.~(\ref{ubt}). One can calculate these using the fact that the state in Eq.~(\ref{ubt}) arises from successive Bogoliubov transformations. These expressions will then be used to calculate the thermal expectation value of the Hamiltonian, and to compute the thermodynamic potential. With $\tilde\psi(\vec{k})$ as defined in Eq.~(\ref{psiexp}), we evaluate the expectation values:
\begin{equation}
\langle \Omega(\beta,\mu)
|\tilde\psi_\alpha^{ia}(\vec{k})\tilde\psi ^{jb}_\beta(\vec{k}')^{\dagger}
|\Omega(\beta,\mu)\rangle
=\delta^{ij}\delta^{ab}
\Lambda_{+\alpha\beta}^{ia}(\vec{k},\beta,\mu)\delta(\vec{k}-\vec{k}'),
\label{psipsidb}
\end{equation}
and
\begin{equation}
\langle \Omega(\beta,\mu)
|\tilde\psi_\beta^{ia\dagger}(\vec{k})\tilde\psi_\alpha^{jb}(\vec{k}')
|\Omega(\beta,\mu)\rangle
=\delta^{ij}\delta^{ab}
\Lambda_{-\alpha\beta}^{ia}(\vec{k},\beta,\mu)\delta(\vec{k}-\vec{k}').
\label{psidpsib}
\end{equation}
Here,
\begin{eqnarray}
\Lambda_\pm^{ia}(\vec{k},\beta,\mu)
&=&\frac{1}{2}\big[1\mp(\sin^2\theta_-- \sin^2\theta_+)\pm \big(\gamma^0\cos\phi_i+\nonumber\\
&&\vec{\alpha}\cdot{\hat{k}}\sin \phi_i \big)\big(1-\sin^2\theta_--\sin^2\theta_+\big)\big].
\label{prpb}
\end{eqnarray}
In Eq.~(\ref{prpb}), we have introduced the  notation $\phi_i(\vec{k})=\phi_i^0(\vec{k})-2 h_i(\vec{k})$ in favor of the condensate function $h(\vec{k})$, which will later prove suitable for variation of the thermodynamic potential.

Using Eqs.~(\ref{psipsidb})-(\ref{prpb}), we can evaluate the expectation value of the NJL Hamiltonian in Eq.~(\ref{ham}) as
\begin{eqnarray}
\epsilon&=&\langle\Omega(\beta,\mu)|{\cal H}|\Omega(\beta,\mu)\rangle\nonumber\\
&=&-\frac{2N_c N_F}{(2\pi)^3}\int d\vec{k} ~ \left[m\cos\phi(\vec{k})+k\sin\phi(\vec{k})\right]\times \left(1-\sin^2\theta_ -- \sin^2\theta_+\right)\nonumber\\
&& -G\left[\left(1+\frac{1}{4N_c}\right)\rho_s^2-\frac{1}{2N_c}\rho_v^2\right],
\label{energy}
\end{eqnarray}
where $N_F$ is the number of flavors. In Eq.~(\ref{energy}), we have set $m_i \equiv m$ and $\phi_i(\vec{k}) \equiv \phi(\vec{k})$. We have also defined the condensates. The scalar condensate is
\begin{equation}
\rho_s=\langle \bar\psi\psi\rangle=-\frac{2N_c N_F}{(2\pi)^3}
\int d\vec{k}\cos\phi(\vec{k})\left(1-\sin^2\theta_--\sin^2\theta_+\right) ,
\label{rhos}
\end{equation}
and the expectation value of the number density is
\begin{equation}
\rho_v=\langle \psi^\dagger\psi\rangle=\frac{2N_c N_F}{(2\pi)^3}
\int d\vec{k}~\left(\sin^2\theta_--\sin^2\theta_+\right) .
\label{rhov}
\end{equation}
The thermodynamic grand potential is then given by
\begin{equation}
\Omega=\epsilon-\mu \rho_v-\frac{1}{\beta}s,
\label{omega}
\end{equation}
where $s$ is the entropy density for the quarks. We have the expression \cite{tfd}
\begin{eqnarray}
s & = & -\frac{2N_c N_F}{(2\pi)^3}\int d\vec{k}~\Big[\sin^2{\theta_-}\ln \left(\sin^2{\theta_-}\right)+\cos^2{\theta_-}\ln\left(\cos^2{\theta_-}\right) \nonumber \\ 
&& \hspace{3cm} +  \sin^2{\theta_+}\ln\left(\sin^2{\theta_+}\right) +\cos^2{\theta_+}\ln \left(\cos^2{\theta_+}\right)\Big ].
\label{ent}
\end{eqnarray}

Now, if we extremize $\Omega$ with respect to $h(\vec{k})$, or equivalently with respect to the function $\phi(\vec{k})$, we obtain
\begin{equation}
\cot \phi(\vec{k}) = \frac{M}{k},
\label{tan2h}
\end{equation}
where $M = m - 2g\rho_s$ with $g=G(1+1/4N_c)$. Substituting this in Eq.~(\ref{rhos}), we have the mass gap equation for the quarks as
\begin{equation}
M = m + 2g \frac{2N_c N_F}{(2\pi)^3}\int d\vec{k}~\frac{M}{\sqrt{k^2 +M^2}} (1-\sin^2\theta_--\sin^2\theta_+).
\label{mgap}
\end{equation}
Similarly, minimization of the thermodynamic potential with respect to the thermal functions $\theta_{\pm}(\vec{k})$ gives
\begin{equation}
\sin^2\theta_\pm = \frac{1}{\exp(\beta\omega_\pm)+1},
\label{them}
\end{equation}
where $\omega_\pm=\sqrt{k^2+M^2}\pm\nu\equiv\epsilon(\vec{k})\pm\nu$. Here, $\nu$ is the interaction-dependent chemical potential given as
\begin{equation}
\nu=\mu-\frac{G}{N_c}\rho_v.
\label{nu}
\end{equation}

In Eq.~(\ref{omega}), we substitute the expression for the condensate function from Eq.~(\ref{tan2h}), and the distribution functions from Eq.~(\ref{them}), to obtain
\begin{eqnarray}
\Omega(M,\beta,\mu)&=&-\frac{12}{(2\pi)^3\beta}\int d\vec{k}~\left\{\ln\left[1+\exp(-\beta\omega_-)\right]+\ln\left[1+\exp(-\beta\omega_+)\right] \right\}
\nonumber\\
&&-\frac{12}{(2\pi)^3}\int d\vec{k}~ \sqrt{k^2+M^2}+g\rho_s^2 
-\frac{G}{6}\rho_v^2,
\label{omegam}
\end{eqnarray}
where we have written down the expression for $N_c=3$ and $N_F=2$. From Eq.~(\ref{omegam}), we subtract the potential of the non-condensed state at $T=0$ and $\mu=0$ to obtain
\begin{eqnarray}
\tilde\Omega(M,\beta,\mu)&=&\Omega(M,\beta,\mu)-\Omega_0(m,\beta=\infty,\mu=0)\nonumber\\
&=&-\frac{12}{(2\pi)^3\beta}\int d\vec{k}~\left\{\ln\left[1+\exp(-\beta\omega_-)\right]+\ln\left[1+\exp(-\beta\omega_+)\right]\right\}
\nonumber\\
&&-\frac{12}{(2\pi)^3}\int d\vec{k}~ \left(\sqrt{k^2+M^2}-\sqrt{k^2+m^2}\right)\nonumber\\
&&+g\rho_s^2-g\rho_{s0}^2 -\frac{G}{2N_c}\rho_v^2.
\label{tomega}
\end{eqnarray}
Here, $\rho_{s0}$ is the scalar density at zero temperature and quark mass $m$:
\begin{equation}
\rho_{s0}=-\frac{12}{(2\pi)^3}\int d\vec{k}\frac{m}{\sqrt{k^2+m^2}}.
\label{rho}
\end{equation}

\subsection{Landau Theory for Chiral Transitions}
\label{glt}

In the mean-field approximation and near the chiral transition line, the thermodynamic potential obtained above can also be described by Landau theory. Let us focus on the case with zero current quark mass. We consider the potential in Eq.~(\ref{tomega}) with $m=0$, and terms of order $N_c^{-1}$ being neglected (i.e., $N_c \rightarrow \infty$):
\begin{align}
\tilde\Omega(M,\beta,\mu) =& \;-\frac{12}{(2\pi)^3\beta}\int \! d\vec{k}\;\Big\{ \ln\left[1+e^{-\beta\left( \sqrt{k^2+M^2} - \mu \right)}\right] \notag\\
& \qquad\qquad\qquad\quad + \ln\left[1+e^{-\beta\left( \sqrt{k^2+M^2} + \mu \right)}\right] \Big\}   \notag\\
&\; -\frac{12}{(2\pi)^3}\int \! d\vec{k} \; \left(\sqrt{k^2+M^2}-k\right)+ \frac{M^2}{4G}.
\label{tomega1}
\end{align}

\begin{figure}[!b]
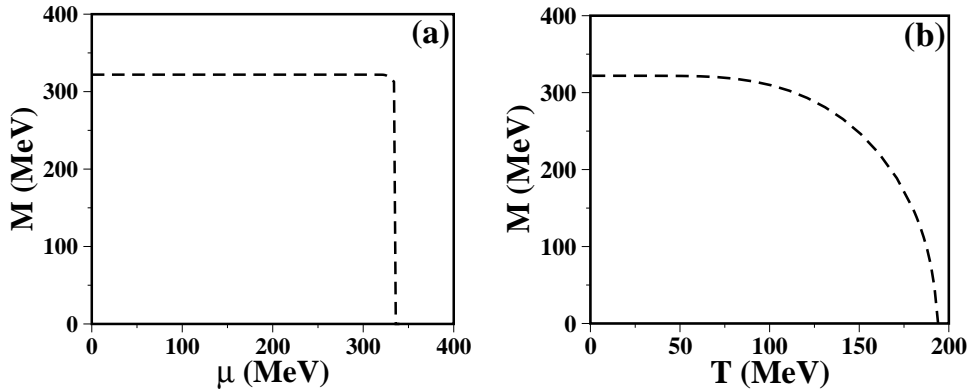

\begin{center}
\begin{tabular}{c c }
\includegraphics[width=0.45\textwidth]{fig1a.eps}&
\includegraphics[width=0.45\textwidth]{fig1b.eps}\\
\end{tabular}
\end{center}
\vspace{-0.6cm}
\caption{(a) Variation of the mass gap $M$ (proportional to the scalar order parameter $\langle\bar\psi\psi\rangle$) with quark chemical potential $\mu$ at $T=0$. (b) Variation of $M$ with $T$ at $\mu = 0$.}
\label{fig1}
\end{figure}

To compute this potential numerically, we set the three-momentum ultraviolet cut-off $\Lambda=653.3$ MeV, and the four-fermion coupling $G=5.0163\times 10^{-6}$ MeV$^{-2}$ \cite{askawa}. With these values, the constituent quark mass at $\mu =0$ and $T=0$ is $M \simeq 312$ MeV. The variation of $M$ with $\mu$ at $T=0$ is shown in Fig.~\ref{fig1}(a). For $\mu < \mu_1(T=0) \simeq 326.321$ MeV, the quark masses stay at their vacuum values. A first-order transition takes place at $\mu = \mu_1$, and the masses of $u$ and $d$ quarks drop from their vacuum values to zero. In Fig.~\ref{fig1}(b), we show the $T$-dependence of $M$ at $\mu=0$. Chiral symmetry is restored for quarks at $T\simeq190$ MeV. In this case, the transition is second-order: this is reflected in the smooth variation of the mass, which is proportional to the order parameter $\langle\bar\psi\psi\rangle$. 

\begin{figure}[!b]
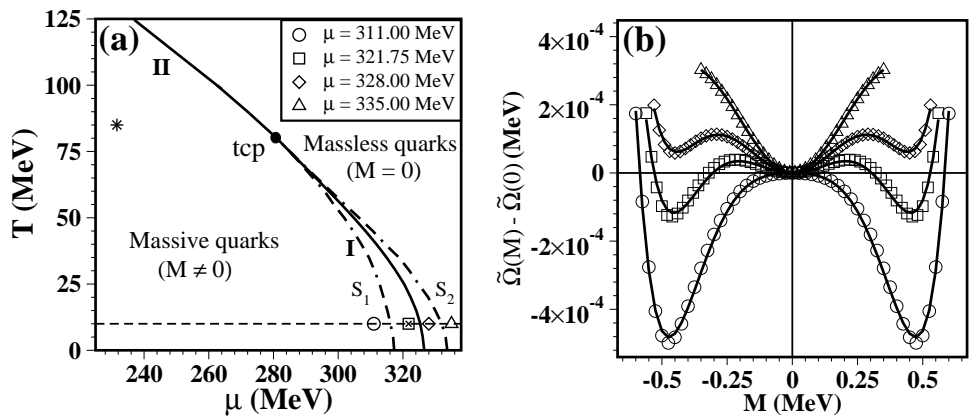

\centering
\begin{tabular}{c c }
\includegraphics[width=0.44\textwidth]{fig2a.eps}&
\includegraphics[width=0.46\textwidth]{fig2b.eps}\\
\end{tabular}
\caption{(a) Phase diagram of the Nambu-Jona-Lasinio (NJL) model in the ($\mu, T$)-plane for zero current quark mass. A line of first-order transitions (I) meets a line of second-order transitions (II) at the tricritical point (tcp). We have $(\mu_\text{tcp}, T_\text{tcp}) \simeq (282.58, 78)$ MeV. The dot-dashed lines $S_1$ and $S_2$ denote the spinodals or metastability limits for the first-order transitions. The open symbols denote 4 combinations of $\left(\mu, T\right)$ with $T=10$ MeV, chosen to represent qualitatively different shapes of the NJL potential. The asterisk and cross denote quench parameters for the simulations described in Sec.~\ref{case1} and Sec.~\ref{case2}, respectively. (b) Plot of $\tilde\Omega\left(M, \beta, \mu \right) - \tilde\Omega\left(0, \beta, \mu \right)$ vs. $M$ from Eq.~(\ref{tomega1}). For a particular $(\mu, T)$-value, we denote the free energy by the same open symbol as in (a). The solid lines superposed on the potentials denote the Landau potential in Eq.~(\ref{p6}) with $a$ from Eq.~(\ref{coff}), and $b$, $d$ being fit parameters (see Table~\ref{tab}).}
\label{fig2}
\end{figure}

In Fig.~\ref{fig2}(a), we show the phase diagram resulting from Eq.~(\ref{tomega1}) for the chiral transition in the ($\mu, T$)-plane. The solid line is the critical line, and corresponds to the chiral phase transition, which can be either first-order or second-order. The first-order line I (at high $\mu$ and low $T$) meets the second-order line II (at low $\mu$ and high $T$) in a tricritical point ($\mu_\text{tcp}$, $T_\text{tcp}$) $\simeq$ ($282.58$, $78.0$) MeV. A first-order transition is characterized by the existence of metastable phases, e.g., supersaturated vapor. The masses corresponding to these metastable phases are local minima of the potential, but have higher free energy than the stable phase. The limit of metastability is denoted by the dot-dashed lines ($S_1$ and $S_2$) in Fig.~\ref{fig2}(a) -- these are referred to as {\it spinodal} lines.

Before proceeding, we should stress that the phase diagram in Fig.~\ref{fig2}(a) only considers homogeneous chiral condensates. However, recent calculations by Carignano et al. \cite{cnb10}, Sadizkowski and Broniowski \cite{sadiz}, and Nakano and Tatsumi \cite{nt05} show the existence of {\it inhomogeneous} chiral-symmetry-breaking phases in the NJL model, e.g., domain-wall solitons, chiral density waves, chiral spirals. In that case, the first-order line (and the associated spinodal lines) in Fig.~\ref{fig2}(a) may be replaced by second-order transitions between inhomogeneous phases. In this paper, we confine ourselves to the kinetics of phase transitions between the homogeneous phases in Fig.~\ref{fig2}(a). However, it is also of great interest to study the ordering dynamics from (say) a homogeneous phase to an inhomogeneous phase, or between different inhomogeneous phases. For example, there have been some studies of ordering to a lamellar (striped) phase in {\it Rayleigh-Benard convection}, described by the {\it Swift-Hohenberg equation} \cite{evg92,cm95,cb98,gk10}. Another important system with ordering to inhomogeneous phases is that of {\it phase-separating diblock copolymers} \cite{bo90,qw96,rh01}.

Close to the phase boundary, the thermodynamic potential (which is even in $M$) may be expanded as a Landau potential in the order parameter $M$: 
\begin{equation}
\tilde\Omega\left(M \right)= \tilde\Omega\left(0 \right) + \frac{a}{2}M^2 + \frac{b}{4}M^4 + \frac{d}{6}M^6 + O(M^8) \equiv f\left(M \right),
\label{p6}
\end{equation}
correct upto logarithmic factors \cite{sasaki,iwasaki}. In the following, we consider the expansion of $\tilde\Omega\left(M \right)$ upto the $M^6$-term. This will be sufficient to recover the phase diagram in Fig.~\ref{fig2}(a), as we see shortly. The first two coefficients [$\tilde\Omega(0)$ and $a$] in Eq.~(\ref{p6}) can be obtained by comparison with Eq.~(\ref{tomega1}) as
\begin{align}
\tilde\Omega(0) =&\;-\dfrac{6}{\pi^2\beta}\displaystyle\int_0^\Lambda \!\!\! dk\,\, k^2 \left\lbrace  
\ln\left[1+e^{-\beta(k-\mu)}\right] + \ln\left[1+e^{-\beta(k+\mu)}\right]\right\rbrace, \nonumber \\
a =& \; \dfrac{1}{2G} - \dfrac{3\Lambda^2}{\pi^2} + \dfrac{6}{\pi^2}\displaystyle\int_0^\Lambda \!\!\! dk\,\,k\left[ \dfrac{1}{1+e^{\beta(k-\mu)}} + \dfrac{1}{1+e^{\beta(k+\mu)}}\right]. 
\label{coff}
\end{align}
We treat the higher coefficients in Eq.~(\ref{p6}) ($b$ and $d$) as phenomenological parameters. These are obtained by fitting $\tilde\Omega\left(M \right)$ in Eq.~(\ref{p6}) to the integral expression for $\tilde\Omega$ in Eq.~(\ref{tomega1}). There are two free parameters in the microscopic theory ($\mu$ and $T$), so we consider the $M^6$-potential with fitting parameters $b$ and $d$. For stability, we require $d>0$.

In Fig.~\ref{fig2}(b), we plot $\tilde\Omega\left(M \right)-\tilde\Omega\left(0\right)$ vs. $M$ from the integral expression in Eq.~(\ref{tomega1}). We show plots for 4 values of $\left(\mu, T\right)$ as marked in Fig.~\ref{fig2}(a). These are chosen to represent qualitatively different shapes of the potential. The solid lines superposed on the data sets in Fig.~\ref{fig2}(b) correspond to the Landau potential in Eq.~(\ref{p6}) with $a$ from Eq.~(\ref{coff}), and $b$, $d$ being fit parameters. The values of these parameters in dimensionless units are provided in Table~\ref{tab}. 
\begin{table}[!ht]
\begin{center}
\renewcommand{\tabcolsep}{0.35cm}
\renewcommand{\arraystretch}{1.5}
    \begin{tabular}{| c | c | c | c | c |}
    \hline
    $(\mu,T)$ (MeV) & $a/\Lambda^2$ & $b$ & $d\Lambda^{2}$ & $\lambda=|a|d/|b|^2$ \\ \hline
    (311.00,10) & -1.306$\times 10^{-3}$ &  0.092 & 0.439 & 0.067 \\ 
    (321.75,10) &  3.539$\times 10^{-3}$ & -0.101 & 0.402 & 0.140 \\ 
    (328.00,10) &  6.431$\times 10^{-3}$ & -0.111 & 0.396 & 0.206 \\
    (335.00,10) &  9.736$\times 10^{-3}$ & -0.101 & 0.265 & 0.255 \\ \hline
    \end{tabular}
\end{center}
\caption{The coefficients ($a,b,d$) of the Landau potential in Eq.~(\ref{p6}) for 4 different values of $\mu$ at $T=10$ MeV. These parameters are specified in dimensionless units of $\Lambda^2$, $\Lambda^0$ and $\Lambda^{-2}$ respectively, where $\Lambda=653.3$ MeV. The dimensionless quantity $\lambda=|a|d/|b|^2$ will be useful in our discussion of the dynamics in Sec.~\ref{kct}.}
\label{tab}
\end{table}

\begin{figure}[!htb]
\centering
\includegraphics[width=0.75\textwidth]{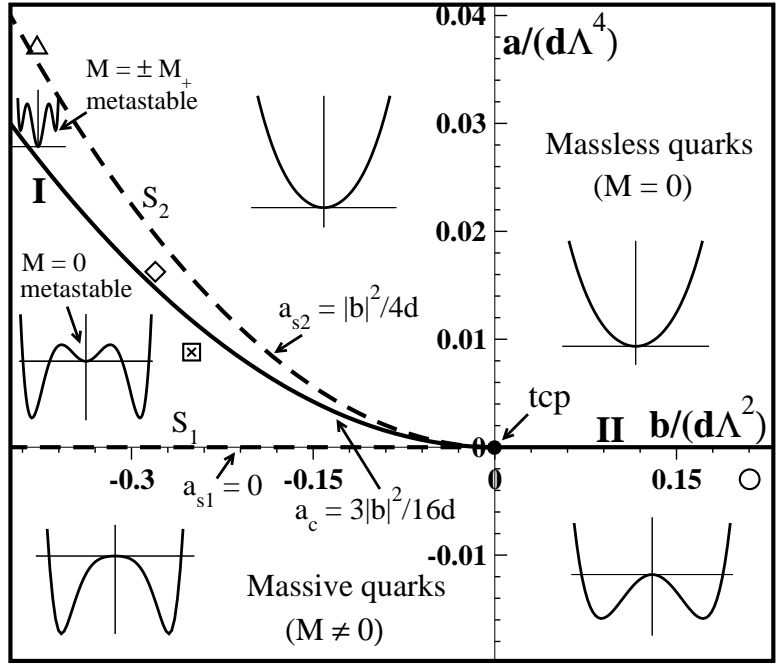}
\caption{Phase diagram for the Landau free energy in Eq.~(\ref{p6}) in
the [$b/(d\Lambda^2), a/(d\Lambda^4)$]-plane. A line of first-order transitions
 (I) meets a line of second-order transitions (II) at the tricritical point (tcp),
 which is located at the origin. The equation for I is $a_c = 3|b|^2/(16d)$,
and that for II is $a_c = 0$. The dashed lines denote the spinodals $S_1$ and $S_2$,
with equations $a_{S_1}=0$ and $a_{S_2} = |b|^2/(4d)$. The typical forms of the
 Landau potential in various regions are shown in the figure. The open symbols
denote the ($\mu,T$)-values marked by the same symbols in Fig.~\ref{fig2}(a). The cross
 denotes the point where we quench the system for $b<0$. The asterisk in Fig.~\ref{fig2}
(a) corresponds to $(a/\Lambda^2,b,d\Lambda^2) = (-1.591 \times 10^{-2}, 8.985
\times 10^{-2}, 7.083 \times 10^{-2})$ or $b/(d\Lambda^2) = 1.269$,
$a/(d\Lambda^4) = -0.225$. We do not mark this point in the figure as
it results in a loss of clarity.}
\label{fig3}
\end{figure}

The order parameter values which extremize the Landau potential are given by the gap equation:
\begin{equation}
f'\left(M \right)= aM + bM^3 + dM^5=0.
\label{ge1}
\end{equation}
The solutions of Eq.~(\ref{ge1}) are
\begin{eqnarray}
M &=& 0, \nonumber \\
M^2 &=& M_{\pm}^2=\dfrac{-b\pm \sqrt{b^2 -4ad}}{2d}.
\label{ges}
\end{eqnarray}
The phase diagram for the Landau potential in $[b/(d\Lambda^2), a/(d\Lambda^4)]$-space is shown in Fig.~\ref{fig3}. For $b>0$, the transition is second-order, as for the $M^4$-potential. The stationary points are $M=0$ (for $a>0$) or $M=0$, $\pm M_+$ (for $a<0$). For $a<0$, the preferred equilibrium state is the one with massive quarks.

For $b<0$, the solutions of the gap equation are
\begin{eqnarray}
M &=& 0, \quad a>|b|^2/(4d), \nonumber \\
M &=& 0, \pm{M_+}, \pm{M_-}, \quad |b|^2/(4d)>a>0, \nonumber \\
M &=& 0, \pm{M_+}, \quad a<0.
\label{mm1}
\end{eqnarray} 
As $a$ is reduced from large values, 5 roots appear at $a=|b|^2/(4d)$. However, this does not correspond to a phase transition. On further reduction of $a$, a first-order transition occurs at $a_c =3|b|^2/(16d)$. The order parameter jumps discontinuously from $M=0$ to $M=\pm M_+$, where $M_+ = [3|b|/(4d)]^{1/2}$. The tricritical point is located at $b_\text{tcp}=0$, $a_\text{tcp}=0$ [cf. Fig.~\ref{fig2}(a)]. The 4 combinations of $\left(\mu, T\right)$-values marked in Fig.~\ref{fig2}(a) are identified using the same symbols in Fig.~\ref{fig3}.

\section{Kinetics of Chiral Transitions}
\label{kct}

\subsection{Dynamical Equation}

Let us next study time-dependent problems in the context of the NJL or Landau free energy.  Consider the dynamical environment of a heavy-ion collision. As long as the evolution is slow compared to the typical re-equilibration time, the order parameter field will be in local equilibrium. We consider a system which is rendered thermodynamically unstable by a rapid quench from the massless phase to the massive phase in Figs.~\ref{fig2}(a) or \ref{fig3}. In the context of Fig.~\ref{fig3}, this corresponds to (say) quenching from $a>a_c(b)$ to $a<a_c(b)$ at a fixed value of $b$. (Of course, we can consider a variety of different quenches.) The unstable massless state (with $M \simeq 0$) evolves via the emergence and growth of domains rich in the preferred massive phase (with $M=\pm M_+$). There has been intense research interest in this far-from-equilibrium evolution \cite{aj94,pw09}. Most problems in this area traditionally arise from materials science and metallurgy. However, equally fascinating problems are associated with the kinetics of phase transitions in high-energy physics or cosmology \cite{tk76,tk07,kp08}.

In this paper, we focus on domain growth in QCD transitions, modeled by the $M^6$-free energy in Eq.~(\ref{p6}). The coarsening system is inhomogeneous, so we include a surface-tension term in the Landau free energy:
\begin{align}
\Omega\left[M\right] =& \; \int \! d\vec{r} \left[f\left(M\right) + \frac{K}{2} \left(\vec{\nabla}M\right)^2\right] \notag \\
=& \; \int \! d\vec{r} \left[\frac{a}{2}M^2 + \frac{b}{4}M^4 + \frac{d}{6}M^6 + \frac{K}{2} \left(\vec{\nabla}M\right)^2\right].
\label{Om}
\end{align}
In Eq.~(\ref{Om}), $\Omega\left[M\right]$ is a functional of the spatially-dependent order parameter field $M(\vec{r})$, and $K$ measures the energy cost of spatial inhomogeneities, i.e., surface tension.

The evolution of the system is described by the {\it time-dependent Ginzburg-Landau} (TDGL) equation:
\begin{align}
\frac{\partial}{\partial t}M\left(\vec{r},t\right)= -\Gamma \frac{\delta \Omega\left[M \right]}{ \delta M}+\theta\left(\vec{r},t\right), 
\label{ke}
\end{align}
which models the over-damped (relaxational) dynamics of $M\left(\vec{r},t\right)$ to the minimum of $\Omega\left[M\right]$, i.e., the system is damped towards the equilibrium configuration \cite{hohenrev}. In Eq.~(\ref{ke}), $\Gamma$ denotes the inverse damping coefficient. The noise term $\theta(\vec{r},t)$ is taken to be Gaussian and white, and satisfies the fluctuation-dissipation relation \cite{hohenrev}:
\begin{align}
\left\langle \theta\left(\vec{r},t\right) \right\rangle =& \; 0,\notag \\
\left\langle \theta(\vec{r'},t')\theta(\vec{r''},t'') \right\rangle =& \; 2\Gamma T\delta(\vec{r'}-\vec{r''})\delta\left(t'-t''\right). 
\label{fdr}
\end{align}
In Eq.~(\ref{fdr}), the angular brackets denote an averaging over different noise realizations. Replacing the potential from Eq.~(\ref{Om}) in Eq.~(\ref{ke}), we obtain
\begin{align}
\frac{\partial}{\partial t}M\left(\vec{r},t\right)= -\Gamma\left(aM + bM^3 + dM^5\right)  + \Gamma K\nabla^2 M + \theta\left(\vec{r},t\right).
\label{ke1}
\end{align}

We use the natural scales of order parameter, space and time to introduce dimensionless variables: 
\begin{eqnarray}
M &=& M_0 M', \quad M_0=\sqrt{|a|/|b|} , \nonumber \\
\vec{r} &=& \xi \vec{r'}, \quad \xi=\sqrt{K/|a|} , \nonumber \\
t &=& t_0 t', \quad t_0=(\Gamma|a|)^{-1} , \nonumber \\
\theta &=& (\Gamma |a|^{3/2}T^{1/2}/|b|^{1/2})~\theta' .
\label{scale}
\end{eqnarray}
Dropping primes, we obtain the dimensionless TDGL equation:
\begin{align}
\frac{\partial}{\partial t}M\left(\vec{r},t\right)= -\mathrm{sgn}\left(a\right)M - \mathrm{sgn}\left(b\right)M^3 - \lambda M^5 + \nabla^2 M +\theta\left(\vec{r},t\right),
\label{ke2}
\end{align}
where $\mathrm{sgn}(x)=x/|x|$ and $\lambda = |a|d/|b|^2 >0$. The values of $\lambda$ corresponding to $T=10$ MeV and $\mu =311,321.75,328,335$ (in MeV) are specified in Table~\ref{tab}. The dimensionless noise term obeys the fluctuation-dissipation relation:
\begin{align}
\left\langle \theta\left(\vec{r},t\right) \right\rangle =& \; 0,\notag \\
\left\langle \theta(\vec{r'},t')\theta(\vec{r''},t'') \right\rangle =& \; 2 \epsilon\; \delta(\vec{r'}-\vec{r''})\delta\left(t'-t''\right),  \notag  \\
\epsilon =& \frac{T |b|}{|a|^{1/2}K^{3/2}}. 
\label{fdr1}
\end{align}

Our results in this paper are presented in dimensionless units of space and time.
 To obtain the corresponding physical units, one has to multiply by the 
appropriate dimensional length-scale $\xi$ and time-scale $t_0$. For this, 
we need to estimate the strength of the interfacial energy $K$, and the
 inverse damping coefficient $\Gamma$. The surface tension can be calculated 
as $\sigma = \sqrt{K}(|a|^{3/2}/|b|)\int dz~(dM/dz)^2$. For quark matter,
 $\sigma$ is poorly known and varies from 10-100 MeV/$\text{fm}^2$ at
 small temperatures \cite{hc93} -- we take $\sigma \simeq 50$ MeV/$\text{fm}^2$.
 For $T=10$ MeV and $\mu = 321.75$ MeV, we then estimate $\xi = \sqrt{K/|a|}
 \simeq 2.8$ fm. Similarly, we set $\Gamma^{-1} \sim 2T/s$, where $s$ is a 
quantity of order 1 \cite{fragaplb,kk92}. This leads to $t_0 = (\Gamma |a|)^{-1} 
\simeq 2.6$ fm/$s$.

We study the phase-transition kinetics for two different quench possibilities. The first case corresponds to high $T$ and low baryon density ($\mu$), where the quenching 
is done through the second-order line (II) in Fig.~\ref{fig2}(a) or Fig.~\ref{fig3}. 
The corresponding parameter values are $(\mu, T) = (231.6, 85)$ MeV 
[marked by an asterisk in Fig.~\ref{fig2}(a)]; or $(a/\Lambda^2,b,d\Lambda^2) = (-1.591
 \times 10^{-2}, 8.985 \times 10^{-2}, 7.083 \times 10^{-2})$ with
 $\lambda = |a|d/|b|^2 = 0.14$.

The second case corresponds to low $T$ and high baryon density ($\mu$), where the chiral dynamics can probe the metastable region of the phase diagram. This can be achieved by shallow quenching through the first-order line (I) in Fig.~\ref{fig2}(a) or Fig.~\ref{fig3}, i.e., quenching to the region between I and S$_1$. This case is studied using parameter values $(\mu, T) = (321.75, 10)$ MeV; or $(a/\Lambda^2, b, d\Lambda^2) = ( 3.53885 \times 10^{-3}, -0.1005344, 0.4015734)$ with $\lambda = 0.14$. These points are marked by a cross in the phase diagrams of Figs.~\ref{fig2}(a) and \ref{fig3}.

\subsection{Quench through Second-order Line: Spinodal Decomposition}
\label{case1}

Let us first focus on the ordering dynamics for quenches through the second-order line ($b > 0$) in Fig.~\ref{fig3}. For $b>0$, the chiral transition occurs when $a<0$. The quenched system is spontaneously unstable and evolves via {\it spinodal decomposition} \cite{aj94,pw09}. The relevant TDGL equation is
\begin{align}
\frac{\partial}{\partial t}M\left(\vec{r},t\right)= M - M^3 - \lambda M^5 + \nabla^2 M +\theta\left(\vec{r},t\right),
\label{ke3}
\end{align}
with the dimensionless potential
\begin{align}
f\left(M \right)= -\frac{1}{2}M^2 + \frac{1}{4}M^4 + \frac{\lambda}{6}M^6.
\label{dp6}
\end{align}
The free-energy minima for this potential are
\begin{align}
M=\pm M_+ = \pm \left(\frac{-1+\sqrt{1+4\lambda}}{2\lambda} \right)^{1/2}.
\label{min_dp6}
\end{align}

We solve Eq.~(\ref{ke3}) with $\lambda = 0.14$ numerically using an Euler-discretization scheme. We implement this on a $d=3$ lattice of size $N^3~(N=256)$, with periodic boundary conditions in all directions. For numerical stability, the discretization mesh sizes must obey the condition
\begin{align}
\Delta t< \frac{2\Delta x^2}{4d + \alpha_1 \Delta x^2},
\label{stc1}
\end{align}
where $\alpha_1 = 4+(1-\sqrt{1+4\lambda})/\lambda$. This condition is obtained from a linear stability analysis of Eq.~(\ref{ke3}) by requiring numerical stability of fluctuations about the stable fixed points in Eq.~(\ref{min_dp6}) \cite{yp87,red88}. For all results shown in this paper, we used the mesh sizes $\Delta x = 1.0$ and $\Delta t = 0.1$. We have confirmed that this spatial mesh size is sufficiently small to resolve the interface region, i.e., the boundary between domains with order parameter $-M_+$ and $M_+$. Further, we use an isotropic approximation to the Laplacian term $\nabla^2M$: 
\begin{align}
\nabla^2M\left(\vec{r}, t\right) = \frac{1}{\Delta x^2} \left[\frac{1}{2} \displaystyle\sum_{\text{nn}}M + \frac{1}{4}\displaystyle\sum_{\text{nnn}}M - 6 M\left(\vec{r}, t\right)\right],
\label{lap}
\end{align}
which couples each cell to its 6 nearest neighbors (nn) and 12 next-nearest neighbors (nnn). Finally, the thermal noise $\theta(\vec{r}, t)$ is mimicked by uniformly-distributed random numbers between $[-A_n, A_n]$. We set $A_n = 0.5$, corresponding to $\epsilon = A_n^2{\left(\Delta x \right)}^d\Delta t/3 = 0.008$ in Eq.~(\ref{fdr1}). This noise amplitude is adequate to initiate the growth process from a metastable state on reasonable time-scales \cite{sp02}, as we will see shortly. However, the asymptotic behavior of domain growth in both the unstable and metastable cases is insensitive to the noise term \cite{yp87,po88}.

\begin{figure}[!htbp]
\centering
\includegraphics[width=0.82\textwidth]{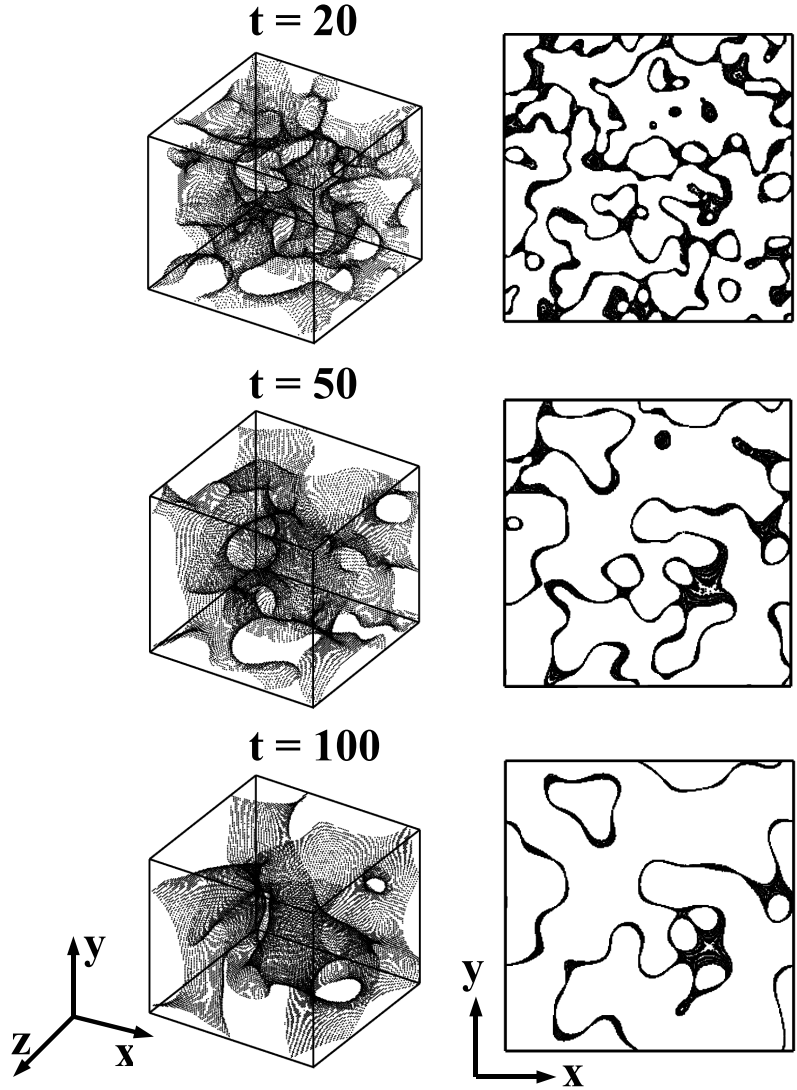}
\caption{Kinetics of chiral transition after a temperature quench through the second-order line (II) in Figs.~\ref{fig2}(a) or \ref{fig3}. The $d$=3 snapshots on the left show the interfaces ($M=0$) at $t=20, 50, 100$ (in units of $t_0$). The defects (interfaces) are kinks between domains of the massive phase with $M\simeq +M_+$ or $M\simeq -M_+$. The snapshots were obtained by numerically solving the TDGL Eq.~(\ref{ke3}) with $\lambda=0.14$, as described in the text. The frames on the right show a cross-section of the snapshots at $z=N/2$.}
\label{fig4}
\end{figure}

In Fig.~\ref{fig4}, we show the evolution of Eq.~(\ref{ke3}) with a disordered initial condition, which consisted of small-amplitude random fluctuations about the massless phase $M=0$. The system rapidly evolves via spinodal decomposition into domains of the massive phase with $M\simeq +M_+$ and $M\simeq -M_+$. The coarsening is driven by interfacial defects, which are shown in Fig.~\ref{fig4}. These domains have a characteristic length scale $L(t)$, which grows with time.

The growth process in Fig.~\ref{fig4} is analogous to coarsening dynamics in the TDGL equation with an $M^4$-potential \cite{aj94,pw09}, i.e., Eq.~(\ref{ke3}) with $\lambda = 0$, which describes coarsening in a ferromagnet subsequent to a temperature quench from $T>T_c$ to $T<T_c$. Coarsening in the ferromagnet is driven by kinks, with the equilibrium profile $M_s(z)= \tanh(\pm z/\sqrt{2})$. We can use the dynamics of kinks to obtain a good understanding of this evolution. The domain scale obeys the Allen-Cahn (AC) growth law, $L(t) \sim t^{1/2}$. Typically, the interface velocity $v\sim dL/dt \sim 1/L$, where $L^{-1}$ measures the local curvature of the interface. This yields the AC law. The same growth law has also been obtained via a closed-time-path formalism of relativistic finite-temperature field theory applied to the NJL model \cite{das}.
 
The pattern morphology in Fig.~\ref{fig4} is statistically self-similar in time with $L(t)$ setting the scale. The morphology is studied experimentally using the {\it order-parameter correlation function} \cite{aj94,pw09}:
\begin{align}
C\left(\vec{r},t\right) = \frac{1}{V}\!\! \int \!\!d\vec{R}\left[ \left\langle M(\vec{R},t)M(\vec{R}+\vec{r},t)\right\rangle - \left\langle M(\vec{R},t)\right\rangle \left\langle M(\vec{R}+\vec{r},t)\right\rangle\right], 
\label{cf} 
\end{align}
or its Fourier transform, the \emph{structure factor}:
\begin{align}
S(\vec{k},t) = \int \!\! d\vec{r}\;e^{i\vec{k}\cdot\vec{r}}C\left(\vec{r},t\right).
\label{sf}
\end{align}
In Eq.~(\ref{cf}), $V$ denotes the volume, and the angular brackets denote an averaging over independent evolutions. As the system is translationally invariant and isotropic, $C(\vec{r},t)$ and $S(\vec{k},t)$ depend only on the vector magnitudes $r$ and $k$. The existence of a characteristic size $L(t)$ results in the {\it dynamical scaling} of $C(\vec{r},t)$ and $S(\vec{k},t)$:
\begin{eqnarray}
C(\vec{r},t) &=& g(r/L), \label{cscale} \\
S(\vec{k},t) &=& L^d f(kL),
\label{dscale}
\end{eqnarray}
where $g(x)$ and $f(p)$ are scaling functions which are independent of time.

\begin{figure}[!htbp]
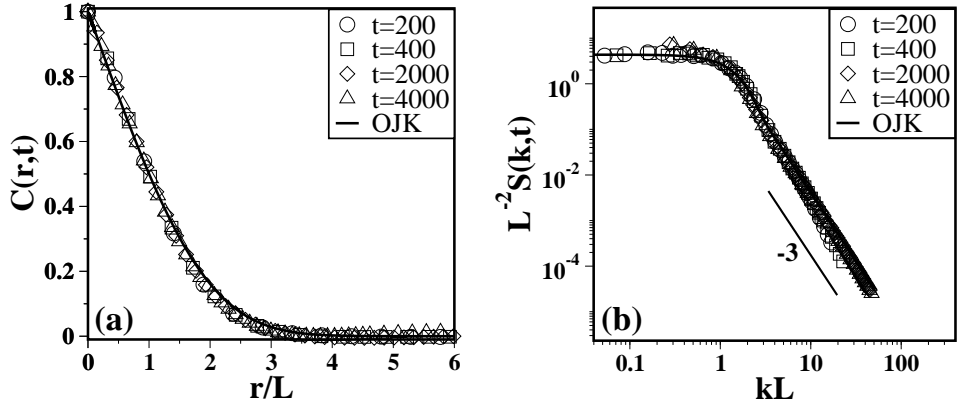

\centering
\begin{tabular}{c c }
\includegraphics[width=0.44\textwidth]{fig5a.eps}&
\includegraphics[width=0.45\textwidth]{fig5b.eps}\\
\end{tabular}
\caption{(a) Dynamical scaling of the correlation function [$C(r,t)$ vs. $r/L$] for chiral spinodal decomposition at four different times. The length scale $L(t)$ is defined as the distance over which $C(r,t)$ decays to half its maximum value. The different data sets collapse onto a single master curve. The statistical data shown in this figure is obtained on $4096^2$ lattices as an average over 10 independent runs, and the correlation function is spherically averaged. The solid line denotes the OJK function in Eq.~(\ref{ojk}) \cite{ojk82}. (b) Dynamical scaling of the structure factor [$L^{-2} S(k,t)$ vs. $kL$] for the same times as in (a). The large-$k$ region (tail) of the structure factor obeys the Porod law \cite{porod}, $S(k,t) \sim k^{-3}$ for $k \rightarrow \infty$, which results from scattering off kink defects.}
\label{fig5}
\end{figure}

Let us demonstrate dynamical scaling for the spinodal decomposition illustrated in Fig.~\ref{fig4}. The statistical results presented in this paper correspond to the $d=2$ case, and are obtained as an average over 10 independent runs with $4096^2$ lattices. In Fig.~\ref{fig5}(a), we plot the scaled correlation function [$C\left(\vec{r},t\right)$ vs. $r/L$] for 4 different times during the evolution. The length scale $L$ is obtained as the distance over which the correlation function decays to half its maximum value [$C(r,t)=1$ at $r =0$]. The data sets collapse onto a single master curve, confirming the scaling form in Eq.~(\ref{cscale}). The solid line in Fig.~\ref{fig5}(a) denotes the analytical result due to Ohta et al. (OJK) \cite{ojk82}, who studied ordering dynamics in a ferromagnet. The magnet is also described by a scalar order parameter, i.e., magnetization. The OJK function is
\begin{align}
C\left(\vec{r},t\right) =& \; \frac{2}{\pi}\sin^{-1}\left(e^{-r^2/L^2}\right).
\label{ojk}
\end{align}
(The corresponding result for the case with vector order parameter has been obtained by Bray and Puri \cite{bp91}.) Our correlation-function data is in excellent agreement with the OJK function, showing that chiral spinodal decomposition is analogous to domain growth in a ferromagnet.

\begin{figure}[!htbp]
\centering
\includegraphics[width=0.47\textwidth]{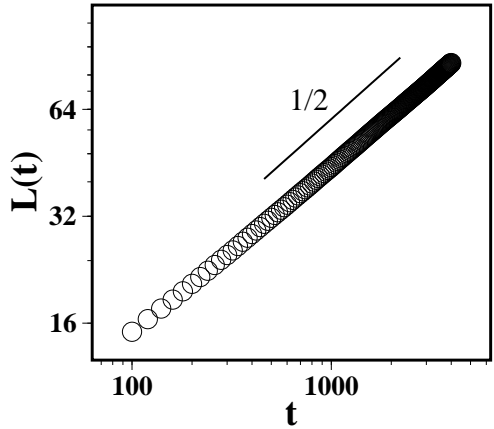}
\caption{Time-dependence of the domain size, $L(t)$ vs. $t$, for chiral spinodal decomposition (see Fig.~\ref{fig5}). The coarsening domains obey the Allen-Cahn (AC) growth law, $L(t)\sim t^{1/2}$.}
\label{fig6}
\end{figure}

In Fig.~\ref{fig5}(b), we plot the scaled structure factor [$L^{-2}S(\vec{k},t)$ vs. $kL$] for the same times as in Fig.~\ref{fig5}(a). Again, the data sets collapse neatly onto a single master curve, confirming the scaling form in Eq.~(\ref{dscale}). The scaling function is in excellent agreement with the corresponding OJK function. Notice that the tail of the structure factor shows the {\it Porod law} \cite{porod}, $S(k,t) \sim k^{-(d+1)}$ for $k \rightarrow \infty$, which results from scattering off sharp interfaces. For an $n$-component vector order parameter, the system shows the {\it generalized Porod law} \cite{bp91}, $S(k,t) \sim k^{-(d+n)}$ for $k \rightarrow \infty$. The scalar order parameter studied here corresponds to $n=1$ and the relevant defects are interfaces or kinks. Higher-order defects arise for vector fields, e.g., vortices or vortex strings ($n=2$), monopoles ($n=3$), etc.

In Fig.~\ref{fig6}, we plot $L(t)$ vs. $t$ on a log-log scale. The length scale data is consistent with the AC growth law, $L(t) \sim t^{1/2}$. As stated earlier, this law has also been obtained from finite-temperature field theory applied to the NJL model \cite{das}.

\subsection{Shallow Quench through First-order Line: Nucleation and Growth}
\label{case2}

Let us next consider shallow quenches from the massless state through the first-order line ($b<0$) in Fig.~\ref{fig3}. The chirally-symmetric phase is now metastable, and evolves to the stable massive phase via the {\it nucleation and growth} of droplets. From Fig.~\ref{fig3}, we see that the first-order chiral transition occurs for $a<a_c=3|b|^2/(16d)$. (In terms of dimensionless variables, the transition occurs for $\lambda<\lambda_c=3/16$.) We consider quenches from $a>a_c$ (with $M=0$) to $a<a_c$. If we quench to $a<0$, the free energy has a double-well structure, as shown in Fig.~\ref{fig3}. Again, the ordering dynamics is analogous to that for the ferromagnet. We have confirmed (results not shown here) that the domain growth scenario is similar to that shown in Figs.~\ref{fig4}-\ref{fig6}.

Subsequently, we focus only on quenches to $0<a<a_c$ or $0<\lambda<\lambda_c$. The appropriate dimensionless TDGL equation is
\begin{align}
\frac{\partial}{\partial t}M\left(\vec{r},t\right)=\; -M + M^3 - \lambda M^5 + \nabla^2 M +\theta\left(\vec{r},t\right) ,
\label{ke4}
\end{align}
with the potential
\begin{align}
f\left(M \right)= \frac{1}{2}M^2 - \frac{1}{4}M^4 + \frac{\lambda}{6}M^6 .
\label{tp6}
\end{align}
The free-energy extrema are located at
\begin{align}
M=\;0, \; \pm M_+,\; \pm M_- ,
\end{align}
where
\begin{align}
M_+ =& \; \left(\frac{1+\sqrt{1-4\lambda}}{2\lambda} \right)^{1/2}, \notag \\
M_- =& \; \left(\frac{1-\sqrt{1-4\lambda}}{2\lambda} \right)^{1/2}.
\end{align}
The extrema at $M=0, \pm M_+$ are local minima with $f(\pm M_+)<f(0)=0$ for $\lambda<\lambda_c$.

\subsubsection{Bubble Growth and Static Kinks}

Before we study the ordering dynamics of Eq.~(\ref{ke4}), it is useful to understand the traveling-wave and static solutions. After all, the phase transition is driven by the dynamics of kinks and anti-kinks (eventually 1-$d$ in nature). For the case with $b>0$ and $a<0$ discussed in Sec.~\ref{case1}, the kinks have tanh-profiles with small corrections due to the $M^6$-term in the potential.

We consider the deterministic version of Eq.~(\ref{ke4}) in $d$=1:
\begin{align}
\frac{\partial}{\partial t}M\left(z,t\right)= \: -M + M^3 - \lambda M^5 + \frac{\partial^2 M}{\partial z^2}.
\label{dke4}
\end{align}
We focus on traveling-wave solutions of this equation, $M\left(z,t\right) \equiv M\left(z-vt\right) \equiv M\left(\eta\right)$ with velocity $v>0$. This reduces Eq.~(\ref{dke4}) to the ordinary differential equation:
\begin{align}
\frac{d^2 M}{d\eta^2} + v\frac{d M}{d\eta} - M + M^3 -\lambda M^5 = 0.
\label{ode}
\end{align}
Equation (\ref{ode}) is equivalent to a 2-$d$ dynamical system:
\begin{align}
\frac{d M}{d\eta} =& \; y, \notag \\
\frac{d y}{d\eta} =& \; M - M^3 + \lambda M^5 -vy.
\label{ode1}
\end{align}
To obtain the kink solutions of this system, we undertake a phase-plane analysis. The {\it phase portrait} will enable us to identify kink solutions of Eq.~(\ref{dke4}).

The relevant fixed points (FPs) are $(M,y)= (0,0)$, $(\pm M_-,0)$, $(\pm M_+,0)$. We consider small fluctuations about these FPs:
\begin{align}
M =& \; M_0 + \phi, \quad (M_0 =0, \; \pm M_-, \; \pm M_+), \notag \\
y =& \; 0+y.
\end{align}
We can linearize Eq.~(\ref{ode1}) about these FPs to obtain
\begin{align}
\frac{d \phi}{d\eta} =& \; y, \notag \\
\frac{d y}{d\eta} =& \; \left(1-3M_0^2+5\lambda M_0^4 \right)\phi -vy \equiv a\phi - vy.
\label{ode2}
\end{align}
The eigenvalues ($\lambda_e$) which determine the growth or decay of these small fluctuations are determined from
\begin{align}
\left|\begin {array}{cc}
-\lambda_e & 1\\
\noalign{\medskip}
a & -v-\lambda_e
\end {array}\right| = 0,
\label{array}
\end{align}
or
\begin{align}
\lambda_{e\pm} = \frac{-v \pm \sqrt{v^2 + 4a}}{2}.
\label{ev}
\end{align}

\begin{figure}[!htbp]
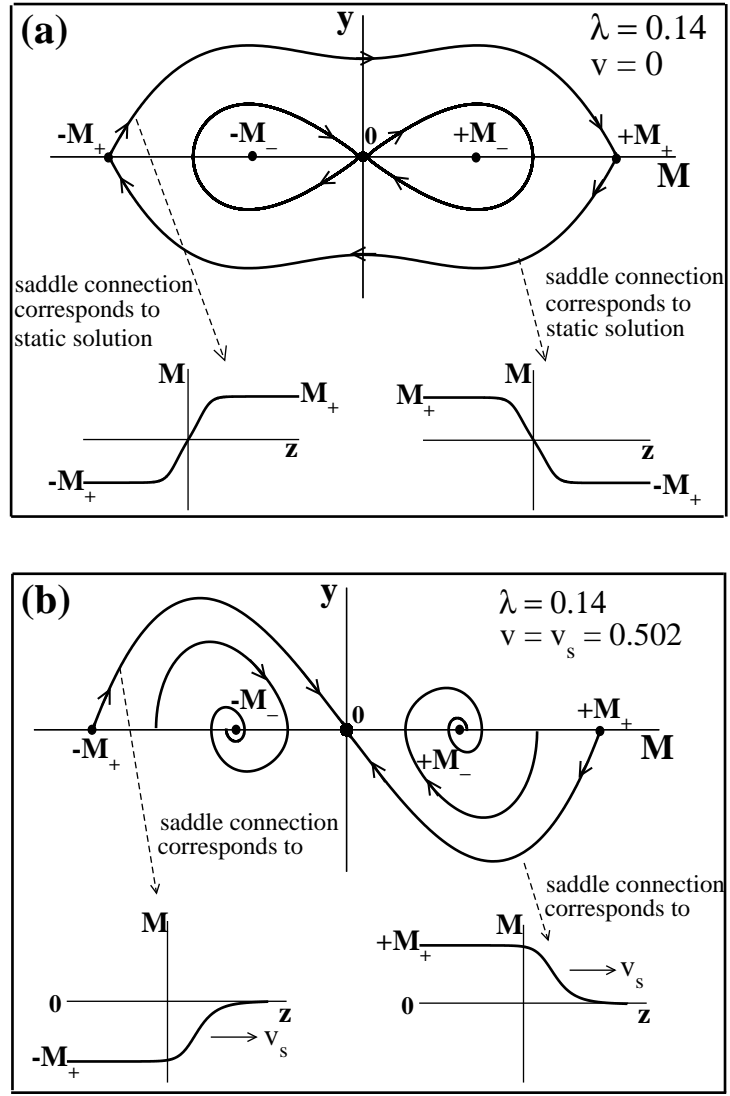

\centering
\includegraphics[width=0.7\textwidth]{fig7a.eps}\\
\vspace{0.6cm}
\includegraphics[width=0.7\textwidth]{fig7b.eps}\\
\caption{Phase portraits of the dynamical system in Eq.~(\ref{ode1}) with
$\lambda = 0.14$. (a) Case with $v = 0$. The saddle connections from
 $-M_+ \rightarrow +M_+$ and $+M_+ \rightarrow -M_+$ correspond to static
kink solutions. (b) Case with $v=v_s = 0.503$, corresponding to the
 appearance of saddle connections from $-M_+ \rightarrow 0$ and
 $+M_+ \rightarrow 0$. These correspond to kinks traveling with velocity $v_s > 0$.}
\label{fig7}
\end{figure}

\begin{figure}[!htbp]
\centering
\includegraphics[width=0.6\textwidth]{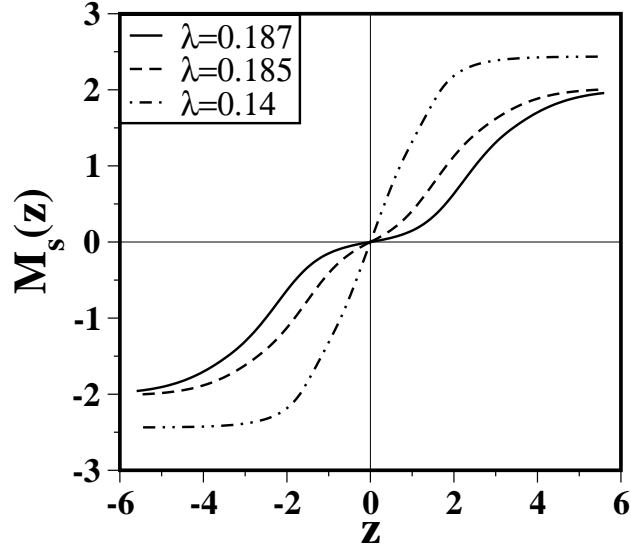}
\caption{Spatial dependence of the order parameter [$M_s(z)$ vs. $z$] for
static kink profiles of Eq.~(\ref{dke4}) with different $\lambda$-values ($< \lambda_c$).}
\label{fig8}
\end{figure}

We can combine this information to obtain the phase portrait of the system in Eq.~(\ref{ode1}). In Fig.~\ref{fig7}, we show phase portraits for $\lambda=0.14$ ($<\lambda_c \simeq 0.1875$), which is the parameter value for most simulations presented in this subsection. Figure~\ref{fig7}(a) corresponds to the case with $v=0$ (static solution). The saddle connections from $-M_+ \rightarrow +M_+$ and $+M_+ \rightarrow -M_+$ correspond to static kink solutions. These can be obtained by integrating
\begin{align}
\frac{d M_s}{dz} = \pm \sqrt{2}\left[\frac{1}{2}\left(M_s^2 - M_+^2 \right) - \frac{1}{4}\left(M_s^4 - M_+^4 \right) + \frac{\lambda}{6}\left(M_s^6 - M_+^6 \right)\right]^{1/2}.
\label{kink1}
\end{align}
The static kink profiles for several values of $\lambda < \lambda_c$ are shown in Fig.~\ref{fig8}.

In Fig.~\ref{fig7}(b), we show the phase portrait for $v = v_s$, where $v_s$ corresponds to the appearance of saddle connections from $-M_+ \rightarrow 0$ and  $+M_+ \rightarrow 0$. These correspond to kinks traveling with velocity $v_s >0$, as shown in Fig.~\ref{fig7}(b). So far, our analysis has been done for the case with $v>0$, but it is straightforward to extend it to the case with $v<0$. In the latter case, the portrait in Fig.~\ref{fig7}(b) is inverted, and the saddle connections (kinks) are from $0 \rightarrow -M_+$ and  $0 \rightarrow +M_+$.

\begin{figure}[!htbp]
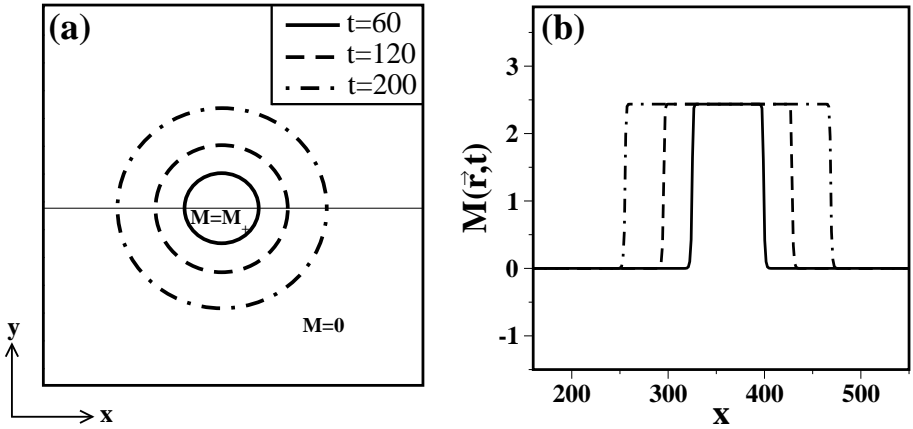

\centering
\begin{tabular}{c c }
\includegraphics[width=0.41\textwidth]{fig9a.eps}&
\includegraphics[width=0.44\textwidth]{fig9b.eps}\\
\end{tabular}
\caption{(a) Growth of a bubble or droplet of the massive phase ($M = +M_+$) in a background of the metastable massless phase ($M=0$). We show the boundary of the droplet at three different times. The innermost circle corresponds to the droplet at time $t=60$. (b) Variation of order parameter along the horizontal cross-section marked in (a).}
\label{fig9}
\end{figure}

\begin{figure}[!htbp]
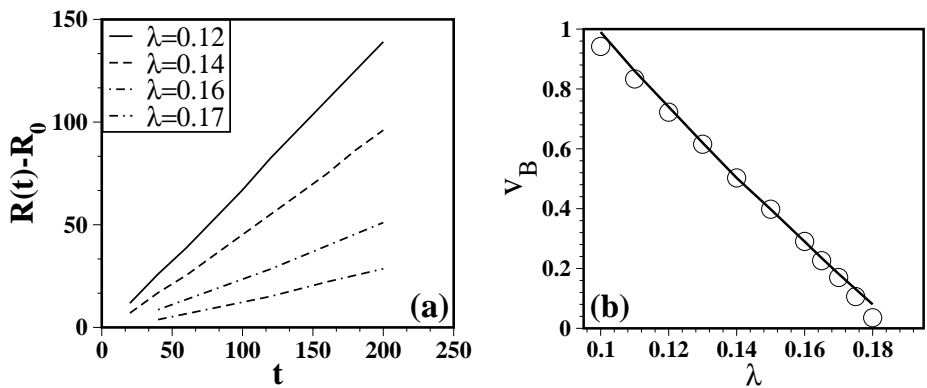

\centering
\begin{tabular}{c c }
\includegraphics[width=0.45\textwidth]{fig10a.eps} &
\includegraphics[width=0.41\textwidth]{fig10b.eps}\\
\end{tabular}
\caption{(a) Linear growth of a single bubble with time [$R(t)-R_0$ vs. $t$] for 4 different values of $\lambda$. (b) Plot of the bubble growth velocity $v_B$ vs. $\lambda$. The circles denote our numerical data, while the solid line is obtained from the phase-plane analysis (see Fig.~\ref{fig7}).}
\label{fig10}
\end{figure}

In Fig.~\ref{fig9}, we show the growth of a bubble (droplet) of the massive phase ($M=+M_+$) in the background of the metastable phase ($M=0$). These snapshots are obtained by solving Eq.~(\ref{ke4}) with $\lambda=0.14$ and $\theta = 0$. We start with an initial configuration of a $d=2$ bubble of radius $R_0>R_c$ such that
\begin{eqnarray}
M(r) &=& M_+, \quad r<R_0 , \nonumber \\
M(r) &=& 0, \quad r>R_0 .
\end{eqnarray}
Here, $R_c(\lambda)$ is the critical size of the droplet, which diverges as $\lambda \rightarrow \lambda_c^-$. In Fig.~\ref{fig10}(a), we plot the radius of the droplet [$R(t) - R_0$] vs. $t$. These curves are linear, showing that the bubble of the massive phase grows at a constant velocity $v_B$. In Fig.~\ref{fig10}(b), we plot $v_B$ vs. $\lambda$. The quantity $(\lambda -\lambda_c)$ measures the degree of undercooling. Our numerical data is in good agreement with $v_s$, which is obtained from the phase-plane analysis [cf. Fig.~\ref{fig7}(b)].

\begin{figure}[!htbp]
\centering
\includegraphics[width=0.7\textwidth]{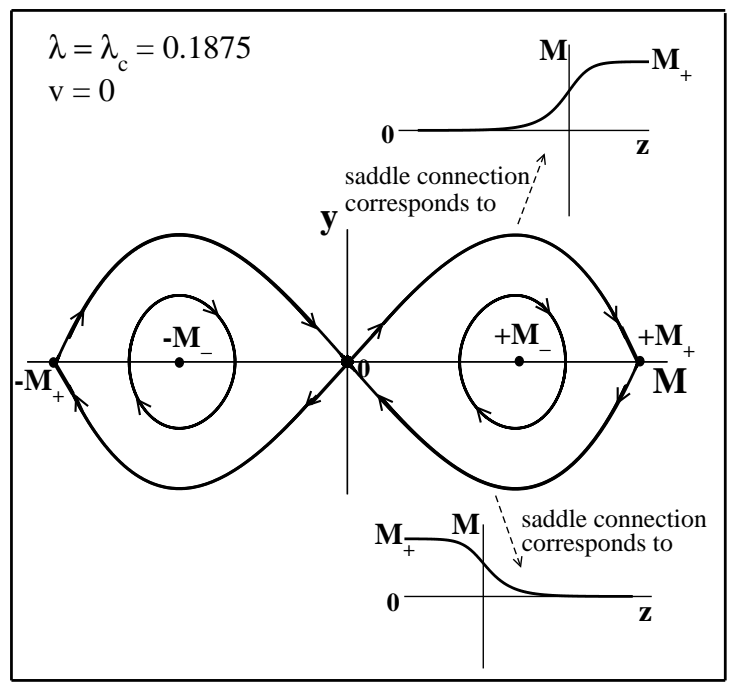}\\
\caption{Phase portrait of the system in Eq.~(\ref{ode1}) with $\lambda = \lambda_c = 0.1875$ and $v=0$. Now, there are saddle connections (or static kinks) from $\pm M_+ \rightarrow 0$ and  $0 \rightarrow \pm M_+$.}
\label{fig11}
\end{figure}

It is also useful to study the limit $\lambda=\lambda_c$, where there are three coexisting solutions of the free energy in Eq.~(\ref{tp6}), viz., $M=0,\; \pm M_+$. The phase portrait for $v=0$ (static kinks) is shown in Fig.~\ref{fig11}. Now, there are saddle connections from $\pm M_+ \rightarrow 0$ and  $0 \rightarrow \pm M_+$. The corresponding kink profiles are obtained as solutions of
\begin{align}
\frac{d M_s}{dz} = \pm \sqrt{2}\left(\frac{1}{2}M_s^2 - \frac{1}{4}M_s^4 + \frac{\lambda_c}{6}M_s^6\right)^{1/2},
\label{kink2}
\end{align}
supplemented with appropriate boundary conditions.


\subsubsection{Chiral Transition Kinetics}

Next, we consider the evolution from a disordered initial condition. We implemented an Euler-discretized version of the TDGL Eq.~(\ref{ke4}) on a $d=2$ lattice of size $N^2$. In this case, the mesh sizes must obey the numerical stability condition:
\begin{align}
\Delta t< \frac{2\Delta x^2}{4d + \alpha_2 \Delta x^2},
\label{stc2}
\end{align}
where $\alpha_2 = -4 + (1+\sqrt{1-4\lambda})/\lambda$. We used the mesh sizes $\Delta x=1.0$ and $\Delta t=0.1$. The other numerical details are the same as in Sec.~\ref{case1}.

\begin{figure}[!htbp]
\centering
\includegraphics[width=0.9\textwidth]{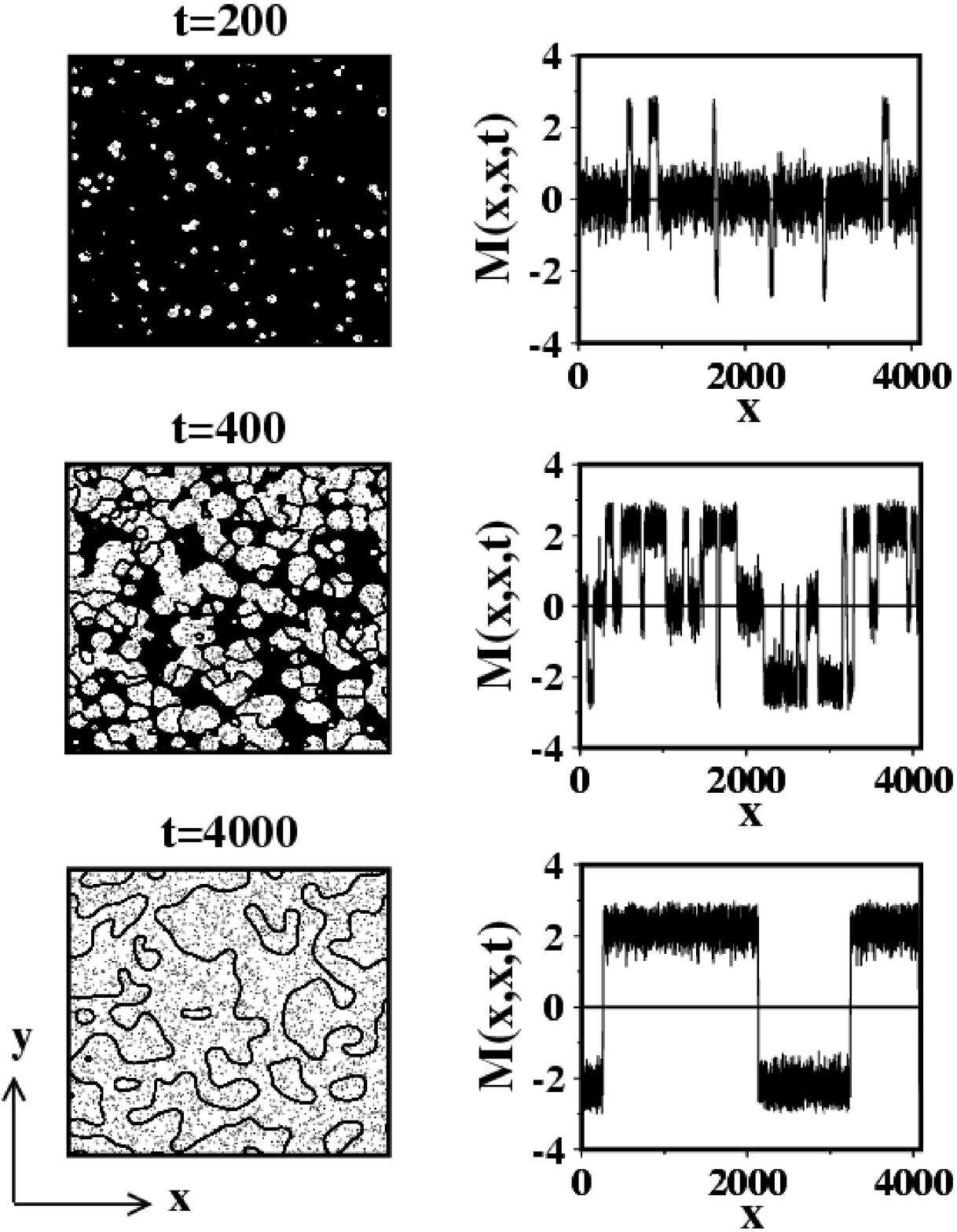}
\caption{Kinetics of chiral transition after a temperature quench through the first-order line (I) in Figs.~\ref{fig2}(a) or \ref{fig3}. The $d=2$ snapshots on the left show regions with $M=0$ at $t=200$, $400$, $4000$ (in units of $t_0$). They were obtained by numerically solving Eq.~(\ref{ke4}) with $\lambda=0.14$. The frames on the right show the variation of the order parameter along a diagonal cross-section ($y=x$) of the snapshots.}
\label{fig12}
\end{figure}

Recall the phase diagram in Fig.~\ref{fig3}. We now focus on the region $b<0$, and consider quenches from high values of $a$ (massless phase) to $0<a<a_c$ or $0<\lambda<\lambda_c$. The massless phase is a metastable state of the potential. The chiral transition proceeds via the nucleation and growth of droplets of the preferred phase ($M=\pm M_+$). This nucleation results from large fluctuations in the initial condition which seed bubbles, or thermal fluctuations during the evolution. This should be contrasted with the evolution in Fig.~\ref{fig4}, where the massless phase is spontaneously unstable and the system evolves via spinodal decomposition. In Fig.~\ref{fig12}, we show the nucleation and growth process which characterizes evolution. At early times ($t=200$), the system is covered with the massless phase, with only small bubbles of the massive phase. The bubbles grow with time (see Fig.~\ref{fig9}) and coalesce into domains ($t=400$). The coarsening of these domains is analogous to that in Fig.~\ref{fig4} -- in the late stages of the transition, there is no memory of the nucleation which enabled growth in the early stages.

\begin{figure}[!htbp]
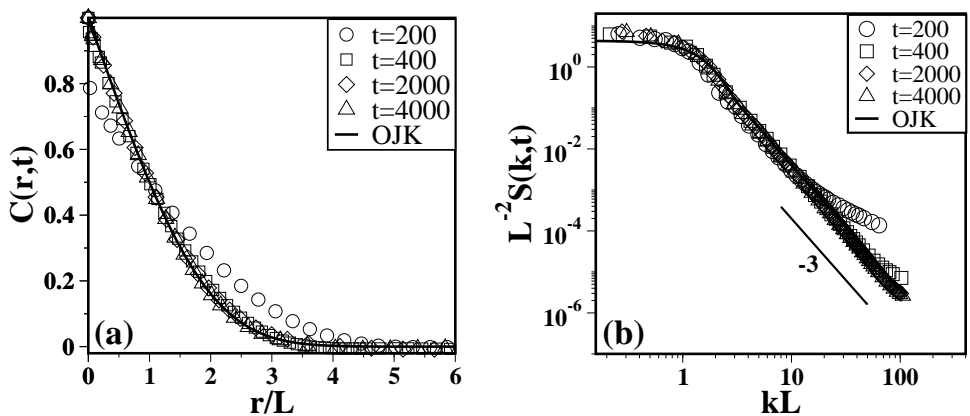

\centering
\begin{tabular}{c c }
\includegraphics[width=0.44\textwidth]{fig13a.eps}&
\includegraphics[width=0.46\textwidth]{fig13b.eps}\\
\end{tabular}
\caption{(a) Scaling plot of the correlation function [$C(r,t)$ vs. $r/L$] for the evolution shown in Fig.~\ref{fig12}. The deviation of data sets from a master curve at early times reflect the morphological differences between the ``nucleation and growth'' and ``domain coarsening'' regimes. At later times, the data sets collapse onto a master curve. The solid line denotes the OJK result in Eq.~(\ref{ojk}). (b) Scaling plot of the structure factor [$L^{-2} S(k,t)$ vs. $kL$] for the same times as in (a). At late times, the tail of the structure factor shows the Porod law, $S(k,t) \sim k^{-3}$ for $k \rightarrow \infty$.}
\label{fig13}
\end{figure}

In Fig.~\ref{fig13}(a), we plot the scaled correlation function [$C(\vec{r},t)$ vs. $r/L$] for the evolution in Fig.~\ref{fig12}. (Our statistical data is obtained as an average over 10 independent runs with $4096^2$ lattices.) The morphological differences between the ``nucleation and growth'' ($t=200,400$) and ``domain coarsening'' ($t=2000,4000$) regimes is reflected in the crossover of the scaling function. At later times, we recover dynamical scaling and the master function is in excellent agreement with the OJK function. Thus, the late-stage morphology in Fig.~\ref{fig12} is analogous to that for spinodal decomposition. In Fig.~\ref{fig13}(b), we plot the scaled structure factor [$L^{-2}S(\vec{k},t)$ vs. $kL$] at the same times as in Fig.~\ref{fig13}(a). As expected, the structure factor also violates dynamical scaling in the crossover regime.

In Fig.~\ref{fig14}(a), we plot the domain size $L(t)$ vs. $t$ for $\lambda = 0.12, 0.14, 0.15$ on a log-log scale. In contrast with Fig.~\ref{fig6}, there is almost no growth in the early stages when droplets are being nucleated. The growth process begins once nucleation is over, with the onset being faster for lower $\lambda$ (or higher degree of undercooling). In Fig.~\ref{fig14}(b), we plot $L(t)-L_0$ vs. $t-t_0$ on a log-log scale. Here, $L_0~(\simeq 10$) is the initial length scale and $t_0$ is the onset time for the different values of $\lambda$ in Fig.~\ref{fig14}(a). We see that the asymptotic regime is again described by the AC growth law, $L(t)\sim (t-t_0)^{1/2}$.

Let us note here that, converting to physical time scales (i.e., multiplying the
dimensionless time $t$ by $t_0$), the time to reach equilibrium seems to be very
large compared to the typical life-time $\tau$ of the fire-ball
produced in a heavy-ion collision ($\tau\sim 10$ fm for the RHIC). Let us recall 
that $t_0 \simeq 2.6$ fm and is proportional to $\Gamma^{-1}$. Larger dissipation 
(larger $\Gamma^{-1}$) will make the equilibration time much larger. Thus, in a
heavy-ion collision experiment, the system may linger in the QGP phase longer during the
fire-ball expansion, even when the temperature has already decreased below the critical
temperature. Thus, the critical temperature calculated within equilibrium thermodynamic
models can be higher than the value that shows up in the growth of fluctuations
in experiments. However, such conclusions depend upon our estimate of $\Gamma$.
In principle, $\Gamma$ can be calculated within the model as has been attempted in
Ref.~\cite{tomoinpa} using the Mori projection operator method.
However, such a calculation is beyond the scope of the present work  and we
have used the estimate of $\Gamma$ in Refs.~\cite{fragaplb,kk92}.

\section{Summary and Discussion}
\label{summary}

Let us conclude this paper with a summary and discussion of the results presented here. We have studied the kinetics of chiral phase transitions in quark matter. At the microscopic level, these transitions are described by the Nambu-Jona-Lasinio (NJL) model. In the NJL model with zero current quark mass, there can be either first-order (I) or second-order (II) transitions between a massless quark phase and a massive quark phase. The lines I and II meet in a tricritical point. At the coarse-grained level, chiral transitions can be modeled by a Landau potential with an $M^6$-functional. We have shown that there is a quantitative agreement between the NJL free energy as a function of $(T,\mu)$ and the Landau potential with an appropriate choice of parameters. Near the phase boundary, we can identify the coefficients of the Landau energy for different values of ($T,\mu$). However, we often consider parameter values far from the phase boundary, and it is more appropriate to interpret the Landau coefficients as phenomenological quantities.

We studied the kinetics of chiral transitions from the massless (disordered) phase to the massive (ordered) phase, resulting from a sudden quench in parameters. We model the kinetics using the {\it time-dependent Ginzburg-Landau} (TDGL) equation, which describes the overdamped relaxation of the order parameter field (scalar condensate density) to the minimum of the corresponding Ginzburg-Landau (GL) free-energy functional. We consider quenches through both 
the first-order and second-order lines in the phase diagram. There have been some earlier studies of the TDGL equation in this context, as discussed in the introductory Sec.~\ref{intro}. However, these have primarily discussed the growth of initial fluctuations in the framework of a linearized theory. On the other hand, this paper focuses on the late stages of pattern formation where nonlinearities in the TDGL equation play an important role.

For quenches through II, the chirally-symmetric phase is spontaneously 
unstable and evolves into the broken-symmetry phase via {\it spinodal decomposition}. 
The evolution morphologies show self-similarity and dynamical scaling, and
can be quantitatively characterized by the {\it order-parameter correlation function} 
or its Fourier transform, the {\it structure factor}. The domains of the massive
phase grow as $L(t) \sim t^{1/2}$.

For deep quenches through I, the above scenario applies again. However, for 
shallow quenches, the chirally-symmetric phase is metastable. Then, the system 
evolves via the {\it nucleation and growth} of bubbles or droplets of the
 preferred massive phase. In this case, the early-stage dynamics is dominated by 
the appearance of bubbles. The growth and merger of these bubbles results in 
late-stage domain growth which is morphologically similar to that for spinodal 
decomposition. The correlation function, structure factor and growth law show a 
crossover, having different functional forms in the nucleation and coarsening 
regimes.

Before concluding, it is important to discuss the relevance of these 
results for QCD phenomenology and experiments. In the context of heavy-ion collisions, 
we make the following observations. Within the uncertainities regarding 
values of dimensional quantities for quark matter (e.g., surface tension, dissipation),
 it is not clear whether the system equilibrates completely within the life-time of the
 fireball. If the system is nearly-equilibrated, the features of the coarsening 
morphology will be similar for quenches through both first- and second-order
 lines in the phase diagram. However, if the equilibration time-scale is
 much larger than the fireball life-time, the morphology is very different for 
quenches through the first-order line, with the system evolving through 
nucleation of bubbles. Consequences of such a first order transition have
 potential relevance since they imply the existence of a critical end point 
(CEP) in the QCD phase diagram. As a matter of fact, experimental studies of such
signatures may be more convenient than directly searching for the CEP
 via critical fluctuations. The latter approach has not provided conclusive
 evidence of the existence of a CEP, presumably due to the smallness of the critical 
region.

We also stress that relating our results to heavy-ion collision experiments 
requires information about the source size apart from its life-time. In this 
context, two-particle momentum correlations (i.e., the {\it Hanbury-Brown-Twiss} 
or HBT effect in heavy-ion collisions) could be relevant. In such two particle
 correlations, the inverse width of the  correlation function in the ``out'' direction 
measures the life-time of the source, 
whereas the same in the ``side'' direction measures the transverse size 
of the source \cite{rischkegyulassy}. Thus, the results presented here regarding 
domain growth and nucleation of bubbles can, in principle, be investigated 
through two-particle HBT correlations. This is similar to
the study of the HBT effect for particle production from inhomogeneous
 clusters of QGP fluid \cite{torrieri}. Finally, the spatial inhomogeneities 
due to the order-parameter evolution could also have measurable effects on the spatial 
distribution and integrated inclusive abundances of various hadrons \cite{dumitru06}.

\vspace{0.5cm}
\noindent {\bf Acknowledgments} \\

HM would like to thank the School of Physical Sciences, Jawaharlal Nehru 
University for hospitality. The authors are also very grateful to the referee
 for constructive and helpful comments. We would like to thank S. Digal and
B. Mohanty for discussions.

\end{document}